\documentstyle[12pt]{article}
\newcommand{\be}{\begin{equation}}
\newcommand{\ee}{\end{equation}}
\newcommand{\rref}[1]{(\ref{#1})}
\begin{document}

\begin{flushright}
ULB--TH--98/02 \\
hep-th/9801053 \\
January 1998\\
\end{flushright}

\vspace{.8cm}

\begin{center}
{\LARGE Statistical Entropy of the Four Dimensional Schwarzschild
Black Hole}
\vspace{2.5cm}

{\large R.~Argurio,}\footnote{
Aspirant F.N.R.S. (Belgium). E-mail:
rargurio@ulb.ac.be}
{\large F.~Englert}\footnote{
E-mail: fenglert@ulb.ac.be}
{\large and L.~Houart}\footnote{
Chercheur I.I.S.N. (Belgium). E-mail: lhouart@ulb.ac.be}
\addtocounter{footnote}{-3}\\
\vspace{.4cm}
{\it Service de Physique Th\'eorique}\\
{\it Universit\'e Libre de Bruxelles, Campus Plaine, C.P.225}\\
{\it Boulevard du Triomphe, B-1050 Bruxelles, Belgium}\\

\end{center}

\vspace{2cm}

\begin{abstract}

The entropy of the four dimensional Schwarzschild black hole 
is derived by mapping it onto a configuration of intersecting
branes with four charges. This configuration is obtained by performing several
boosts and dualities on a neutral black brane of M-theory to which 
the Schwarzschild black hole is related by trivial compactification.
The infinite boost limit is well-defined and 
corresponds to extremality where the intersecting
brane configuration is a marginal one on which a standard 
microscopic counting of the entropy can be safely performed. The result
reproduces exactly the Bekenstein-Hawking entropy of the four dimensional
black hole.

\end{abstract}

\newpage

Recent progress in string theory and M-theory indicates that an 
understanding of fundamental issues in black hole physics may well
be within reach.
Using D-brane techniques a statistical explanation of the
entropy for some black holes has been discovered. The entropy has
been computed in terms of the degeneracy of
D-brane configurations describing in the weak coupling, charged 
black holes in the extremal and near-extremal limit 
\cite{stromvafa,callanmalda,horstrom,maldastrom,johnkhuri,horolow,bala}.
Unfortunately, this systematic approach cannot be applied directly to
neutral Schwarzschild black holes. Microscopic considerations
based on Matrix theory
have however been discussed recently \cite{bfksa,klesu,li,homa,bfksb,ohtazhou}.
Other considerations \cite{skend} involve a connection between the
Schwarzschild black hole and the 2+1 dimensional BTZ black hole \cite{btz},
which has been given a microscopic description in \cite{carlip,strom,birm}.

A quantitative analysis of Schwarzschild black holes applying some M-theory 
concepts has been suggested in \cite{fran}.
It is proposed, following the idea of \cite{das}, to view a
Schwarzschild black hole as a compactification of a black brane
in 11 dimensional supergravity and to relate it to a charged black
hole with the same thermodynamic entropy. The charged black hole is
obtained by subjecting the black brane to a boost \cite{tseyt} in 
uncompactified spacetime followed by Kaluza-Klein reduction
on a different radius \cite{das} (see also \cite{homa}).
In \cite{fran}, a near extremal limit is
defined, in which the Schwarzschild radius remains arbitrarily large
at infinite boost. It is proposed to use this limit to obtain the entropy of
Schwarzschild black holes from the microscopic entropy of the charged ones,
viewed as systems of D-branes. 
This was applied to the seven dimensional black hole, mapped onto
a near extremal system of D3-branes.

In this letter we apply this proposal to four dimensional Schwarzschild
black holes.
In order to relate a four dimensional Schwarzschild black hole to a
``countable" D-brane configuration, the procedure is more involved
because we will have to perform several boosts and dualities \cite{das}. More
precisely, each boost creates a (Ramond-Ramond) charge, and we will end
up with a configuration of intersecting branes with four charges. 
In this case, the infinite boost limit leads exactly to extremality
where the configuration is marginal and a standard microscopic
counting of the entropy can be safely performed (because it is protected
by BPS arguments). In this way
the Bekenstein-Hawking entropy of the four dimensional Schwarzschild 
black hole is exactly recovered.

The metric of a four dimensional Schwarzschild black hole is:
\be
ds^2=-f dt^2+f^{-1} dr^2+r^2 d\Omega_2^2, \qquad f=1-{r_0\over r}.
\label{schwarz} \ee
This metric can be trivially embedded in an 11 dimensional space-time
simply by taking its product with a flat 7-dimensional (compact) space.
It thus corresponds to a neutral black seven-brane compactified on
a $T^6 \times S^1$ characterized by sizes $L_i$ ($i=1\dots 6$) and $2\pi R$.
The four and eleven dimensional Newton constants, respectively
$G_4$ and $G_{11}\equiv l_p^9$, are related by:
\be 
G_4={l_p^9 \over 2\pi L_1\dots L_6 R} \label{newton}.
\ee
The Bekenstein-Hawking entropy of the black hole described by \rref{schwarz}
is given by:
\be 
S_{BH}={\pi r_0^2\over G_4}=2\pi^2 {L_1\dots L_6 R \over l_p^9} r_0^2.
\label{sbh} \ee

We now consider this neutral black seven-brane in the framework of M-theory.
Most generally, we recall the precise relation between the parameters
of M-theory compactified on a circle $S^1$ (i.e. the eleven dimensional
Planck length $l_p$ and the radius $R$) and the parameters of type IIA
string theory, i.e. the string coupling $g_s$ and the string length
$l_s\equiv \sqrt{\alpha'}$:
\begin{eqnarray}
R&=&g_s l_s, \label{r11} \\
l_p^3&=& 2^{4/3} \pi^{7/3} g_s l_s^3. \label{lpl11}
\end{eqnarray}
This gives the 10 dimensional Newton coupling constant:
\be 
G_{10}={l_p^9 \over 2\pi R}=8\pi^6 g_s^2 l_s^8. \label{G10}
\ee

Using boosts (in a sense to be defined below) and dualities, we will
map the above black brane onto a configuration of intersecting branes
carrying 4 Ramond-Ramond charges. 
We will then show that there exists a limit in which the latter
configuration approaches extremality in such a way that the 
statistical evaluation of its entropy is well-defined.
 
We now proceed to the careful description of all the steps leading to the
final configuration corresponding to the intersection
D4$\cap$D4$\cap$D4$\cap$D0, which is a marginal bound state in the
extremal limit.

Let $z$ be a coordinate parametrizing the covering
space of the $S^1$ factor of the compact space
over which the neutral black brane is wrapped. 
We can now perform a boost of rapidity $\alpha$ in that direction.
In the boosted frame, the length $R$ is rescaled to the value:
\be R'={R\over \cosh \alpha}. \label{rprime} \ee
We now define a new compactification identifying the boosted coordinate $z'$
on intervals of length $2\pi R'$. This is not a constant time compactification
as viewed from the unboosted $z$ frame. The 10 dimensional theories defined 
respectively by compactification on $R$ and $R'$ are thus different.
They nevertheless coincide at the horizon where all time intervals
are blueshifted to zero. The two compactifications are thus on the horizon
related by a coordinate transformation, as discussed in \cite{fran}.

The Schwarzschild black hole embedded in 11 dimensions is of course
not invariant under the boost. The metric acquires a non-vanishing
off-diagonal $g_{t'z'}$ component.
The resulting configuration, as viewed in 10 dimensions, is a non-extremal
set of D0$_1$-branes (smeared on the $T^6$); the subscript indicates
that the D0-branes have been created by the first boost. 
This IIA string theory is
characterized by parameters $g_s$ and $l_s$ which have a well-defined
dependence on the boost parameter $\alpha$ given
by \rref{r11} and \rref{lpl11} when $R$ is replaced by $R'$.

In the following step we perform T-dualities in the directions, say,
$\hat{1} \hat{2} \hat{3} \hat{4}$ of $T^6$ to obtain a non-extremal
configuration of D4$_1$-branes. We use the standard T-duality relations:
\be
L_i \rightarrow {4\pi^2 l_s^2 \over L_i}, \qquad
g_s \rightarrow g_s {2\pi l_s \over L_i}. \label{tduality}
\ee

We now uplift this IIA configuration to 11 dimensions. Note that this
is a ``new" M-theory in the sense that the Planck length is now a function
of $\alpha$ and the dependence on $\alpha$ of the radius
of compactification has changed.

To create a second charge, we perform a boost of parameter
$\beta$ on the eleventh direction. Following the same procedure
as for the first boost, the radius of compactification of the new
M-theory is rescaled
by $1/\cosh\beta$. After compactification, the resulting IIA configuration
corresponds to the non-extremal version of the marginal bound state 
D4$_1\cap$D0$_2$.

We then T-dualize on the $\hat{1} \hat{2} \hat{5} \hat{6}$ directions,
leading to a D4$_1\cap$D4$_2$ configuration. The first set of D4-branes
lies now in the $\hat{3} \hat{4} \hat{5} \hat{6}$ directions.

Uplifting to eleven dimension  for the second time, the parameters
now depend on the two boosts $\alpha$ and $\beta$. We are now ready
to create a third charge, performing a third boost of parameter
$\gamma$. 

This results, after compactification and T-dualities over 
$\hat{1} \hat{2} \hat{3} \hat{4}$, to a non-extremal configuration
D4$_1\cap$D4$_2\cap$D4$_3$, lying respectively in the
$\hat{1} \hat{2} \hat{5} \hat{6}$, $\hat{3} \hat{4} \hat{5} \hat{6}$
and $\hat{1} \hat{2} \hat{3} \hat{4}$ directions.

A last uplift--boost--compactification procedure characterized
by a boost parameter $\delta$ leads to our final configuration
D4$_1\cap$D4$_2\cap$D4$_3\cap$D0$_4$. The corresponding metric in the
Einstein frame is (see e.g. \cite{cvetse,aref,nobu}):
\begin{eqnarray}
ds^2&=&-H_{\alpha}^{-{3\over 8}}H_{\beta}^{-{3\over 8}}
H_{\gamma}^{-{3\over 8}}H_{\delta}^{-{7\over 8}}f dt^2+
H_{\alpha}^{-{3\over 8}}H_{\beta}^{5\over 8}
H_{\gamma}^{-{3\over 8}}H_{\delta}^{1\over 8}(dy_1^2+dy_2^2) \nonumber \\
&&+H_{\alpha}^{5\over 8}H_{\beta}^{-{3\over 8}}
H_{\gamma}^{-{3\over 8}}H_{\delta}^{1\over 8}(dy_3^2+dy_4^2) +
H_{\alpha}^{-{3\over 8}}H_{\beta}^{-{3\over 8}}
H_{\gamma}^{5\over 8}H_{\delta}^{1\over 8}(dy_5^2+dy_6^2) \nonumber \\
&&+H_{\alpha}^{5\over 8}H_{\beta}^{5\over 8}
H_{\gamma}^{5\over 8}H_{\delta}^{1\over 8}(f^{-1}dr^2+r^2d\Omega_2^2)
\label{conf}
\end{eqnarray}
where
\be
f=1-{r_0\over r}, \qquad H_\alpha=1+{r_0\over r} \sinh^2 \alpha
\label{harmfns} \ee
and similarly for $H_{\beta}$, $H_{\gamma}$ and $H_{\delta}$. The non-trivial
components of the RR field strengths are:
\begin{eqnarray}
\tilde{F}_{ty_1y_2y_5y_6r}&=&-\partial_r \left(H_\alpha^{-1} {r_0\over r}
\cosh\alpha \sinh\alpha\right), \nonumber \\
\tilde{F}_{ty_3y_4y_5y_6r}&=&-\partial_r \left(H_\beta^{-1} {r_0\over r}
\cosh\beta \sinh\beta\right),  \\
\tilde{F}_{ty_1y_2y_3y_4r}&=&-\partial_r \left(H_\gamma^{-1} {r_0\over r}
\cosh\gamma \sinh\gamma\right), \nonumber \\
F_{tr}&=&-\partial_r\left(H_\delta^{-1} {r_0\over r}
\cosh\delta \sinh\delta\right), \nonumber 
\label{fields}
\end{eqnarray}
where $\tilde{F}_6$ is the 10 dimensional Hodge dual of the 4-form
RR field strength.

The string coupling and string length of the type $\widehat{
\mbox{IIA}}$ theory in which
this configuration is embedded are, in terms of the original quantities
appearing in \rref{newton} and \rref{sbh}:
\begin{eqnarray}
\hat{g}_s&=&4\pi^{1\over2} {l_p^{9\over2}\over L_1L_2L_5L_6 R^{1\over2}}
\left({\cosh\alpha \cosh\beta \cosh\gamma\over \cosh^3\delta}\right)^{1\over2},
\label{ghat}\\
\hat{l}_s^2&=&{1\over 2^{4\over3}\pi^{7\over3}} {l_p^3\over R}
\cosh\alpha \cosh\beta \cosh\gamma \cosh\delta \label{lshat}
\quad (=l_s^2 \cosh\alpha \dots \cosh\delta) 
\end{eqnarray}
The lengths of the final 6-torus over which the above configuration
is wrapped are:
\begin{eqnarray}
\hat{L}_{1,2}&=&{2^{2\over3}\over\pi^{1\over3}}{l_p^3\over R L_{1,2}}
\cosh\alpha \cosh\gamma, \nonumber \\ 
\hat{L}_{3,4}&=&L_{3,4}\cosh\beta \cosh\gamma, \label{Lhat}\\
\hat{L}_{5,6}&=&{2^{2\over3}\over\pi^{1\over3}}{l_p^3\over R L_{5,6}}
\cosh\alpha \cosh\beta. \nonumber 
\end{eqnarray}

We now compute the charge densities of the D-branes in this configuration:
\begin{eqnarray}
Q_{D4_1}&=& {1 \over 16 \pi \hat{G}_{10}} \int_{\hat{T}^2_{(3,4)} \times
S^2} F_4 = {\pi^{7\over3} \over 2^{11 \over 3}} 
{L_1^2 L_2^2 L_3 L_4 L_5^2 L_6^2 R^5 r_0 \tanh \alpha\over l_p^{21} 
\cosh^3 \alpha \cosh^3 \beta \cosh^3 \gamma \cosh \delta}, \nonumber \\
Q_{D4_2}&=& {1 \over 16 \pi \hat{G}_{10}} \int_{\hat{T}^2_{(1,2)} \times
S^2} F_4 = {\pi^{5\over3} \over 2^{7 \over 3}} 
{L_1 L_2  L_5^2 L_6^2 R^3 r_0 \tanh \beta \over l_p^{15}
\cosh^3 \alpha \cosh^3 \beta \cosh^3 \gamma \cosh \delta}, \nonumber \\
Q_{D4_3}&=& {1 \over 16 \pi \hat{G}_{10}} \int_{\hat{T}^2_{(5,6)} \times
S^2} F_4 = {\pi^{5\over3} \over 2^{7 \over 3}} 
{L_1^2 L_2^2  L_5 L_6 R^3 r_0 \tanh \gamma \over l_p^{15} 
\cosh^3 \alpha \cosh^3 \beta \cosh^3 \gamma \cosh \delta}, 
\label{charges} \\
Q_{D0_4}&=& {1 \over 16 \pi \hat{G}_{10}} \int_{\hat{T}^6 \times
S^2}\star F_2 = {\pi \over 2}
{L_1 \dots L_6 R r_0 \sinh \delta \over l_p^{9}
\cosh \alpha \cosh \beta \cosh \gamma }. \nonumber
\end{eqnarray}
The charge densities above are normalized in such a way that
the elementary D-branes have a charge density equal to their 
tension \cite{tasi}.

The tensions of elementary D0 and D4-branes are given by:
\begin{eqnarray}
T_{D0} &=& {1 \over \hat{g}_s \hat{l}_s} = {\pi^{2\over3} \over 2^{4 \over 3}}
{L_1 L_2  L_5 L_6 R \over l_p^{6}} {\cosh \delta
\over \cosh \alpha \cosh \beta \cosh \gamma } \label{tensions}\\
T_{D4} &=& {1 \over 2^4 \pi^4  \hat{g}_s \hat{l}_s^5} = 
{\pi^{4\over3} \over 2^{8 \over 3}}{L_1 L_2  L_5 L_6 R^3 \over l_p^{12}}
{1 \over \cosh^3 \alpha \cosh^3 \beta \cosh^3 \gamma \cosh \delta} \nonumber
\end{eqnarray}

Using \rref{charges} and \rref{tensions}, we can now compute the different
numbers of constituent D-branes of each type:
\begin{eqnarray}
N_1&=&Q_{D4_1} T_{D4}^{-1} = {\pi \over 2}{L_1 \dots L_6 R^2 \over l_p^{9}}
r_0 \tanh \alpha \nonumber \\
N_2&=&Q_{D4_2} T_{D4}^{-1} =( 2\pi)^{1 \over 3}
{L_5  L_6 \over l_p^{3}} 
r_0 \tanh \beta \nonumber \\
N_3&=&Q_{D4_3} T_{D4}^{-1} =( 2\pi)^{1 \over 3}
{L_1  L_2 \over l_p^{3}} 
r_0 \tanh \gamma \label{numbers} \\
N_4&=&Q_{D0_4} T_{D0}^{-1} =( 2\pi)^{1 \over 3}
{L_3  L_4 \over l_p^{3}} 
r_0 \tanh \delta \nonumber 
\end{eqnarray}

Strictly speaking,
these numbers represent the number of branes only in the extremal limit,
but can be interpreted more generally as the difference between the number
of branes and anti-branes.
Note that it is also
possible to compute the numbers \rref{numbers} by evaluating after every boost 
the number of D0 branes created.
Indeed, the numbers are strictly invariant under all the subsequent
dualities and further boosts.

We will now show that taking all the boost parameters to infinity 
with $r_0$ kept fixed is
equivalent to taking the extremal limit on the intersecting D-brane
configuration. In order to do this, we compute the ADM mass:
\be 
M={\pi\over4}{L_1\dots L_6 R\over l_p^9} r_0
{\cosh 2\alpha+\cosh 2\beta +\cosh 2\gamma +\cosh 2\delta\over
\cosh \alpha \cosh \beta \cosh \gamma \cosh \delta}. \label{mass}
\ee
This formula is to be compared with the extremal value derived from the
charge densities \rref{charges}:
\be
M_{ext}=Q_{D4_1}\hat{L}_1\hat{L}_2\hat{L}_5\hat{L}_6+
Q_{D4_2}\hat{L}_3\hat{L}_4\hat{L}_5\hat{L}_6+
Q_{D4_3}\hat{L}_1\hat{L}_2\hat{L}_3\hat{L}_4+
Q_{D0_4}. \label{mext}
\ee
It is then easy to show that when all the boost parameters are equal and
are taken to infinity, the departure from
extremality rapidly goes to zero as:
\be
{M-M_{ext}\over M_{ext}}\sim e^{-4\alpha} \label{extremality}
\ee
Note that in the same limit both $M$ and $M_{ext}$ go to zero
as $e^{-2\alpha}$. However we have to take into account the formulas
\rref{ghat} and \rref{lshat} which tell us that $\hat{g}_s$ remains
finite and $\hat{m}_s\equiv \hat{l}_s^{-1}$ goes to zero as
$e^{-2\alpha}$. Thus the masses above are finite in string units.
Note also that for all the internal directions the ratio
$\hat{L}_i/\hat{l}_s$ is finite\footnote{This result and the
finiteness of $\hat{g}_s$ imply that the string units and the
Planck units in 4 dimensions are of the same order in $\alpha$.}.

Despite the fact that $\hat{m}_s$ is vanishingly small in the limit
discussed, we can neglect the massive string modes because the
Hawking temperature goes to zero even faster. Indeed, we have:
\be
T_H={1 \over 4\pi r_0 \cosh \alpha \cosh \beta \cosh \gamma \cosh \delta}
\label{hawking}
\ee
which gives a $T_H/\hat{m}_s$ of order $e^{-2\alpha}$ in the extremal limit.

For the supergravity description to be valid and not corrected
by higher order curvature terms, we have to check that the typical
length scale derived from the Riemann tensor is much bigger than
the string length. This can indeed be computed, and it holds as soon
as $r_0\gg l_p$. 

In this infinite boost limit, the numbers of constituent D-branes 
\rref{numbers} tend to a finite value, which moreover is large if
we take $r_0\gg l_p$ as above and we assume that $R$ and all $L_i$
are of order $l_p$ or bigger.
The fact that the $N$'s are large but tend to a finite value
is a crucial element for the validity of our mapping procedure.

We are now in position to compute the microscopic degeneracy of this
extremal D4$\cap$D4$\cap$D4$\cap$D0 configuration. This configuration
can indeed be related to the one considered in \cite{maldastrom} by 
a series of T- and S-dualities. One ends up with a configuration
consisting of $N_2$ D6-branes wrapped on the whole $T^6$, $N_3$ 
NS5-branes lying in the $\hat{1} \hat{2} \hat{3} \hat{4} \hat{5}$ directions,
$N_1$ D2-branes in the $\hat{5} \hat{6}$ directions (which become actually
$N_1 N_3$ after breaking on the NS5-branes) and finally $N_4$ quanta of
momentum in the $\hat{5}$ direction. Going to flat space, the degeneracy
of these momentum excitations can be computed as in \cite{maldastrom}.
The statistical entropy is then given for large $N$'s by:
\be
S_{micro}=2\pi \sqrt{N_1 N_2 N_3 N_4}. \label{smicro}
\ee
Note that all the $N$'s, which by \rref{numbers} are proportional to 
$r_0$, can be taken arbitrarily large because our limit has been 
defined keeping $r_0$ fixed and, in fact, arbitrarily large.
Using the values \rref{numbers} in the infinite boost limit,
we find:
\be
S_{micro}=2\pi^2 {L_1\dots L_6 R \over l_p^9} r_0^2= {\pi r_0^2\over G_4},
\ee
in perfect agreement with the Bekenstein-Hawking entropy of the
original four dimensional Schwarzschild black hole given
in \rref{sbh}.

This result calls for some comments.

At first sight, counting states of a Schwarzschild black hole through
a mapping onto an extreme BPS black hole seems a very indirect procedure.
However the mapping crucially rests on the relation between 
compactification radii (see e.g. Eq. \rref{rprime}) 
ensuring equality of semi-classical thermodynamic
entropies. As we have seen, this mapping is equivalent,
on the horizon only, to a coordinate transformation 
in eleven dimensions.
The physics outside the horizon is different and in spacetime we have
two distinct physical systems. Nevertheless, the very fact that entropy
was obtained by a counting of quantum states strongly suggests that the
two different systems can be related by a reshuffling of degrees of
freedom defined on the horizon.

Another crucial element of our computation is the fast convergence
in the infinite boost limit to a configuration of extremal BPS D-branes.
This is in sharp contradistinction to the case examined in \cite{fran}
where all the entropy came from a departure from extremality. This 
arose because the slow vanishing of the excess mass $\Delta M\equiv M-M_{ext}$
was exactly compensated by the growth of the internal volume to give 
a finite value to the entropy of the non-BPS excitations of the D3-branes.
In our case the situation is different. The product $\Delta M L$ goes
to zero in the limit, leaving a pure BPS state which can then be safely
extrapolated to flat space.

\end{document}